# Directed Self-Guided Learning of Blended Math-Science Sensemaking for Historically Marginalized STEM Learners

Leonora Kaldaras and Carl Wieman

## Abstract

**Background**

Blended math-science sensemaking (MSS) is reflected in a student's ability to integrate math and science knowledge to develop mathematical descriptions of observations. While an important component of scientific thinking, there is little research on teaching MSS. Students from backgrounds historically marginalized in STEM often lack prior learning opportunities needed to succeed in STEM. Supporting them in developing MSS could help them build quantitative understanding and improve STEM performance. Fostering MSS calls for supporting transferable, lifelong skills. We designed and tested a self-guided learning model grounded in simulations and a cognitive framework to achieve transferable MSS.

**Methods**

We tested the activities grounded in the framework with 27 students from backgrounds historically marginalized in STEM attending minority serving 2- and 4-year US colleges. We evaluated long and short-term MSS transfer by asking student to complete transfer tasks immediately following and at least 1 week post activity.

**Findings**

Most students started at the lowest and progressed to the highest MSS level and demonstrated considerable short and long-term transfer after completing just one activity, suggesting the effectiveness of this approach for fostering transferable MSS.

**Contribution**

This work provides a roadmap for designing self-guided MSS learning experiences across STEM contexts for a diverse range of learners.

## Introduction

Blended Math-Science Sensemaking (MSS) is reflected in the ability to explain scientific phenomena mathematically by integrating scientific and mathematical reasoning (Redish, 2017; Zhao & Schuchardt, 2021; Kuo et al., 2013). It is through this blending of math and science ideas that we can develop highly precise and predictive models of the natural world. MSS is a foundational cognitive process that lies at the heart of scientific thinking and is indicative of deep science understanding (Redish, 2017; Zhao & Schuchardt, 2021). MSS is also an important component of computational thinking (CT) - one of the fundamental skills emphasized throughout K-16 level (NGSS, National Research Council [NRC], 2013; STEM Education Act, 2015).

While the value of MSS has been well recognized, traditional undergraduate STEM courses don't effectively support the development of this skill (Becker & Towns, 2012; Bing & Redish, 2007; Taasoobshirazi & Glynn, 2009; Tuminaro & Redish, 2007). Most students don't naturally engage in MSS and need instructional support for developing this ability (Kaldaras & Wieman, 2023a, b). Supporting learners in developing MSS skills in introductory STEM courses may help marginalized learners improve their likelihood of successfully completing these courses and benefit all students. Prior research shows that performance in introductory college STEM courses impacts long-term career pathways and is strongly associated with STEM degree completion (Hatfield et al., 2022; Dika & D'Amiko, 2016). Therefore, supporting marginalized learners in developing MSS skills as part of the introductory STEM coursework might increase their likelihood of successfully completing STEM-related programs and pursuing STEM careers.

One of the major functions of education is development of life-long learning skills (Zimmerman, 2000). This is especially important as society seeks to deal with many challenges which require creative solutions. Design of these solutions is closely related the ability of our STEM graduates to constantly develop new skills and make connections across subject matter. This ability, in turn, requires a considerable self-guidance on the learner's part. In the current study, as an intermediate supportive step towards developing true self-guidance, we are using a learning approach of "directed self-guidance" (Brydges et al., 2009), defined as self-guided learning which is informed and structured by external influences.

In the current study we aim to develop and test an instructional approach for directed self-guided learning of MSS with the goal of helping learners develop transferable MSS skills and provide them with the necessary strategies for continually using MSS across STEM settings. We build on our prior work of developing and validating a cognitive model (framework) describing proficiency in MSS across STEM disciplines (Kaldaras & Wieman, 2023a; Kaldaras & Wieman, in preparation) and designing and testing an instructional model for teaching MSS in first-year undergraduate STEM courses aligned to this model (Kaldaras & Wieman, 2023b). The current study will address the following research questions (RQ):

1) *Is the directed self-guided learning model designed in this study effective in helping learners from backgrounds historically marginalized in STEM in developing transferable MSS skills?*

2) *What degree of transfer of MSS skills do learners demonstrate after completing one directed self-guided learning activity focused on MSS?*

To answer RQ1, we designed a directed self-guided learning model grounded in prior work and tested this model by developing activities aligned to the model. We evaluated students' MSS proficiency upon completing the activities to determine the effectiveness of these activities in helping learners develop MSS skills. Further, each activity incorporated a transfer task focused on measuring student MSS skills beyond the STEM context covered by the activity. Students completed the transfer task immediately following activity completion. Students also completed a delayed MSS assessment sometime after activity completion (between 1- 4 weeks). The MSS assessment probes MSS skills across different STEM contexts. Student performance on both the transfer task and the MSS assessment was evaluated to determine the degree of transfer of MSS skills learned in the self-guided activity, which helped answer RQ 2.

## Theoretical Framework

In this paper we draw on a range of theoretical foundations from developmental psychology to design a model for directed self-guided learning of MSS skills with emphasis on transfer. First, we build on our prior work of developing and validating a cognitive framework for MSS (Kaldaras & Wieman, 2023a) and use this framework as a guide for designing and structuring activity tasks. Second, we build on our prior work of developing and testing an instructional model for teaching MSS by leveraging PhET interactive computer simulations (Kaldaras & Wieman, 2023b). We discovered that PhET simulation foster authentic engagement in the dynamic MSS process by allowing learners to explore various quantitative relationships related to the phenomenon under study and offer valuable supports for learners from backgrounds historically marginalized in STEM when they engage in MSS (Kaldaras & Wieman, in preparation). In this study, we are leveraging the power of PhET simulations to design activities that foster directed self-guidance by combining LP-driven cognitively appropriate tasks with technology enhanced feedback and scaffolding embedded in PhET. The combination of these resources offers the necessary external structure, which fosters directed self-guidance- a cognitive process anchored in theories of self-guided learning. Finally, we leverage previous work on knowledge transfer (Schwartz & Bransford, 1998) to incorporate supports that foster transferability of MSS skills across STEM contexts. We will further discuss each of these theoretical anchors in more detail and introduce the learning model for directed self-guided learning of MSS skills that we use in this study.

**Using Cognitive Models to Describe the Developmental Process of Learning**

Supporting learners in developing a deep understanding of a cognitive construct requires time and cognitively appropriate instructional supports to foster progression towards higher levels of understanding (Duschl, 2019). Research shows that cognitive development does not occur in specific age-defined stages, and it is not uniform across individuals (National Research Council [NRC], 2006). Rather, developmentally appropriateness is not a simple function of grade or age, but largely dependent on prior opportunities to learn (NRC, 2006; Kaldaras & Krajcik, 2004). Cognitive models, such as learning progressions (LPs) outline increasingly sophisticated ways of thinking about a construct (NRC, 2006). In accordance with the developmental approach, LPs do not assume that all students progress in a single linear, defined set of levels (Shepard, 2018). Rather, LPs aim to describe the most frequent orderings of ways in which students can think about a topic grounded in empirical research and in the logic of the discipline (Shepard, 2018).

LPs embody a developmental vision of student understanding by allowing accurate diagnosis of the current level of understanding of individual learners and by providing a guide to educators on how to support learners in transitioning to higher levels (Kaldaras & Krajcik, 2024). Therefore, learning progressions shift the goal of the learning process from focusing on what students should be learning and doing based on their age or some other external measure of cognitive development to focusing on creating learning opportunities for students based on their individual level of understanding, which is contingent upon their prior learning opportunities. By enabling this shift, LPs represent a powerful tool for helping to adjust instruction to the needs of individual learners (Kaldaras & Krajcik, 2024).

In this study we are using a previously validated cognitive model for MSS (Kaldaras & Wieman, 2023a) as a framework for designing cognitively appropriate tasks and guiding learners towards higher level MSS during the self-guided activity. We will further discuss the MSS framework in more detail.

*Development and Validation of Cognitive Framework for MSS*

While MSS have been described for specific disciplines (Bing & Redish, 2007; Hunter et al., 2021; Schuchardt & Schunn, 2016), our recent work has formulated and tested a cognitive theory (framework) of MSS applicable across STEM fields (Kaldaras & Wieman, 2023a). The MSS framework shown in Table 1 describes increasingly sophisticated ways of engaging in MSS as students are developing a mathematical equation to describe their observations of a physical phenomenon or building a deeper understanding of a known equation.

The framework consists of three broad levels: qualitative (level 1), quantitative (level 2), and conceptual (level 3). The difference among the three levels lies primarily in ***the degree to which students demonstrate a quantitative account*** of the phenomenon in question, usually in terms of an equation, and ground their account in observations or available evidence. Specifically, level 1 (Qualitative) is reflected in providing mostly qualitative description of the phenomenon without providing a quantitative description. Level 2 (Quantitative) is broadly characterized by providing a quantitative account (i.e., an accurate mathematical equation) describing the phenomenon that is grounded in specific numerical data related to the phenomenon without generalizing to a broader set of observations. Level 3 (Conceptual) is broadly described by providing quantitative account (an accurate mathematical equation) describing the phenomenon that is grounded in quantitative patterns (e.g., proportional relationships) derived from data and observations, thereby reflecting the student's ability to generalize observations beyond a specific set of numerical values to a conceptual relationship.

In addition, each broad level is also divided into three sub-levels titled "Description", "Pattern" and "Mechanism" to reflect different ***types* of MSS** that students can demonstrate at a given broad MSS level. "Description" type of MSS relates to recognizing the variables relevant for describing a given phenomenon mathematically. "Pattern" type of MSS relates to recognizing patterns among the variables that are relevant for describing the phenomenon mathematically. "Mechanism" type of MSS relates to providing causal account of the phenomenon by relating relevant math and science ideas. The degree to which each of these three MSS types reflects quantitative accounts of the phenomenon determines whether the given MSS type characterizes Levels 1, 2 or 3 of the frameworks. To characterize proficiency in MSS according to the framework levels, one needs to determine the broad level of MSS (qualitative, quantitative, conceptual) as well as type of MSS being demonstrated (Description, Pattern, Mechanism) for a given student. Together these two characteristics determine sub-level of the framework.

**Tackling the power of representation for MSS**

While cognitive models can guide the development of cognitively appropriate learning supports, these supports can only serve as effective learning opportunities if student can meaningfully engage with them. Creating feasible opportunities to learn for diverse learners is challenging because it is not enough to simply expose learners to certain information or tasks (Gee, 2008). Rather, it is important to focus on the relationship between the learner and the learning environment and consider whether the learning environment affords the types of learning opportunities that a given group of individuals can perceive as valuable to help them engage in the learning process (Gee, 2008).

MSS is a *dynamic* cognitive process that involves continuous accumulation of evidence associated with changing the parameters of the system and revising an explanation based on new evidence to figure something out (Odden & Russ, 2019). This dynamic nature of the MSS process makes computer simulations such as PhET simulations a promising tool for engaging students in MSS. By limiting cognitive load and providing intuitive visual representations with a minimum of text to interpret, the simulations are effective for students with a wide range of educational backgrounds. They provide students with the freedom to manipulate various parameters, while focusing their exploration only on the aspects that are the most relevant for understanding how relevant variables affect the observations of the phenomena (Adams, 2010; Adams et al., 2015). They represent a promising way of using the power of representation, by offering a simplified (but not too simplified) system for exploring all the relevant mathematical aspects of the system and relating them to the overarching phenomenon, without the need for prior knowledge. In short, PhET simulations offer a tool for helping students from a wide range of backgrounds authentically engage in MSS by lowering the cognitive load, minimizing the requirement of prior knowledge and language proficiency, and offering accessible representations.

**Directed Self-Guided Learning and Knowledge Transfer**

Self-regulation in learning is reflected in self-awareness, self-motivation, and behavioral skill to use the knowledge appropriately (Zimmerman, 2002). As learner progresses from novice to expert, they are developing the ability to recognize when to apply the knowledge they have learned to achieve desired outcomes, which is an indicative feature of self-regulated learner (Zimmerman, 2002) and is also reflective of knowledge transfer ability (Bransford, 2000). Therefore, ability to successfully transfer knowledge to unfamiliar contexts is an important feature of self-regulated learners (Zimmerman, 2002). Research tells us that self-regulation of learning is not a single personal trait that individual students either possess or lack. Instead, it involves the selective use of specific skills that must be personally adapted to each learning task (Clearly & Zimmerman, 2000). This reflects a high degree of generalization of one's understanding, which is an important part of the ability to transfer skills and knowledge to explain a wide range of phenomena (Bransford, 2000).

*Figure 1. Visual representation of differences between SDL, SGL and DSGL.*

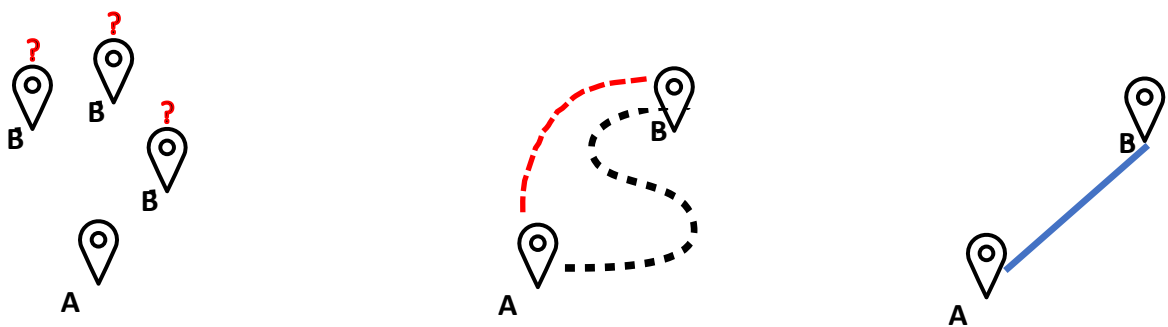

**Self-Directed Learning (SDL):** learner decides what skills and knowledge are needed and what the learning outcome should be given their prior knowledge (A).

**Self-Guided Learning (SGL):** learner decides how to use knowledge and skills gained during the learning process to attain a given learning outcomes (B) given their prior knowledge (A). Alternative approaches for knowledge application that a learner might choose are shown by red and black dotted lines.

**Directed Self-Guided Learning (DSGL):** learner is supported in using the knowledge and skills in specific ways to attain learning outcome (B) using the external structure provided by the learning environment (blue solid line). Learner decides how to use the gained knowledge and skills in alternative ways

Self-Directed learning (SDL) is closely related to self-regulated learning (SRL) but represents a broader concept than SRL. Specifically, SDL assumes that students have control over what will be learned (Loyens et al., 2008). In the current study, we are focusing on fostering the process of self-regulated learning (also sometimes referred to as self-guided learning) as a step towards educating self-directed learners. We recognize that students need considerable supports to develop skills necessary to become a successful self-regulated learners and so we focus on a more restrictive process called directed self-guided learning (Brydges et. al., 2009). Directed self-guided learning (DSGL) constrains the process of self-guided learning via external factors to direct the learner towards productive outcomes (Brydges et al., 2009). We believe it represents a cognitively appropriate form of self-guided learning at the early stages of educating a self-regulated learner because the external structure provides a certain degree of autonomy while directing the learner towards productive learning outcomes. This approach is also consistent with constructivist learning theories emphasizing that guided discovery (or scaffolded inquiry) is often more productive for learners than pure discovery learning reflected in lack of any types of scaffolds (Schwartz & Bransford, 1998). In the context of guided discovery, it is important to let students work on discovering important aspects of the topic under study prior to telling them about the relevant content- a practice that depends on determining the right time for telling (Schwartz & Bransford, 1998). In the approach introduced in this study we follow this practice by letting students explore the phenomenon and answer cognitively appropriate questions before providing them with short text-based summaries that help students generalize their activity and provide important information for meaningful engagement in subsequent guided discovery. This reflects directed self-guided nature of our learning model. We demonstrate the relationship between SRL, SDL and DSGL in figure 1. We will further elaborate on how directed self-guidance is achieved in our proposed learning model.

Table 1. *Blended Math-Sci Sensemaking Framework (sample responses provided in the context of exploring acceleration with PhET simulation shown in Figure 1).*

| | | |
|---|---|---|
| **1 (Qualitative)** | Description | Students use observations to identify measurable quantities (variables) contributing to the phenomenon.<br>*Example: the mass of the box and the force we apply to push the box affect the motion.* |
| | Pattern | Students recognize patterns among the variables identified using observations and explain *qualitatively* how the change in one variable affects other variables, and how these changes relate to the phenomenon.<br>*Example: the lighter object speeds up more than heavier ones when the same force is exerted on both.* |
| | Mechanism | Students demonstrate *qualitative* understanding of the underlying causal scientific mechanism (cause-effect relationships) behind the phenomenon based on the observations but don't define the exact math relationship.<br><u>*Example*</u>: *the same force would cause lighter object to speed up faster compared to heavier objects.* |
| **2 (Quantitative)** | Description | Students recognize that the variables identified using the observations provide measures of scientific characteristics and explain *quantitatively* how the change in one variable affects other variables (but not recognizing the quantitative patterns yet), and how this change relates to the phenomenon in question. Students do not yet to express the phenomenon as an equation.<br><u>*Example*</u>: *when I apply the force of 450 N on a box with mass 50 kg, the acceleration is about 7.12 m/s².* |
| | Pattern | Students *recognize quantitative patterns* among variables and explain *quantitatively (in terms of an equation or formula)* how the change in one parameter affects other parameters, and how these changes relate to the phenomenon in question. Students do not yet relate the observed patterns to the operations in a mathematical equation and don't develop the exact mathematical relationship.<br><u>*Example*</u>: *when applied force changes by 1-unit, acceleration changes by 1 unit, which is a linear relationship.* |
| | Mechanism | Students explain *quantitatively* (express relationship as an equation) how the change in one variable affects other variables based on the numerical values of specific variables. Students include the relevant variables that are not obvious or directly observable. Students do not yet explain conceptually why each variable should be in the equation beyond noting that the specific numerical values of variables and observed quantities match with this equation. Students don't explain how the mathematical operations used in the equation relate to the phenomenon, and why a certain mathematical operation was used beyond the logic of plugging in the numbers (e.g., the equation makes sense because when I plug in the associated values, it always gives me the correct outcome). Students provide qualitative causal account for the phenomenon.<br><u>*Example*</u>: *The formula ($F_{applied} - value$) $= m \times a$ makes sense based on the data. When I apply a force of 100 N to push an* |
| | | *object of 50 kg, the acceleration would be 0.12 m/s². When I apply for the force of 200 N to the same object, acceleration would be 2.12 m/s², so acceleration increases with increasing applied force. This suggests mass should be multiplied by acceleration because if it were divided or subtracted, acceleration would not increase. And I don't think addition would give the correct values. I also notice that you need to subtract something from applied force in order for the equation to work. You need to do (100 N- 94) = 50 kg × 0.12 m/s², and then both sides equal 6, so the equation works. Similarly, when we have 200 N of applied force, we need to do (200 N – value) = 50 kg × 2.12 m/s², so the value would also be 94. Looking at the data, these values correspond to the friction force. So, to make the equation work, you need to do ($F_{applied}$ – friction) =m × a, which makes sense based on the data.* |
| **3 (Conceptual)** | Description | Students describe the observed phenomenon in terms of an equation, and they explain why all variables or constants (including unobservable or not directly obvious ones) should be included in the equation.  Students do not yet explain how the mathematical operations used in the formula relate to the phenomenon.<br>*Example: In F=m × a, the F is always less than applied force by specific number, so there must be another variable subtracted from $F_{applied}$ to make the equation work. The variable involves the properties of the surface because I can see that all the force that I apply does not go into accelerating the object-some goes to overcoming the resistance of the surface. So, the equation should be modified: $F_{applied}$-(friction)=m × a.* |
| | Pattern | Students describe the phenomenon in terms of an equation, and they can explain why all variables or constants (including unobservable or not directly obvious ones) should be included in the equation.  The justification for the equation includes relating quantitative patterns among the relevant variables to the equation structure (such as relating directly proportional relationships to multiplication or inversely proportional relationships to division to explain the relationship) Students do not yet provide a causal explanation of the equation structure.<br>*Example: In $F_{net}$=m × a, multiplication makes sense because as applied force on the same mass increases, acceleration increases linearly, which suggests multiplication.* |
| | Mechanism | Students describe the observed phenomenon in terms of an equation and explain why all variables or constants (including unobservable or not directly obvious ones) should be included in the equation.  Students fully explain how the mathematical operations used in the equation relate to the phenomenon in questions and therefore demonstrate *quantitative* conceptual understanding.  Here, the conceptual understanding is how mathematical relationships represent numerical dependencies.<br>*Example:  greater acceleration is caused by applying a larger net force to a given mass. Therefore, acceleration is a dependent variable. Acceleration is directly related to net force and inversely related to mass. Therefore, the net force should be in the numerator and the mass should be in the denominator such that Acceleration = $F_{net}$/mass.* |

**Learning Model for Directed Self-Guided Learning of MSS**

We developed a directed self-guided learning model for supporting MSS. The model is grounded in previously validated MSS cognitive framework (Table 1), PhET interactive simulations and a combination of feedback and additional scaffolds in the form of short text-based summaries. The model is shown in Figure 2. Note that this is a general model, and variations in tasks, nature of feedback and information provided in summaries might vary depending on the context.

*Figure 2. Directed Self-Guided Learning Model for MSS.*

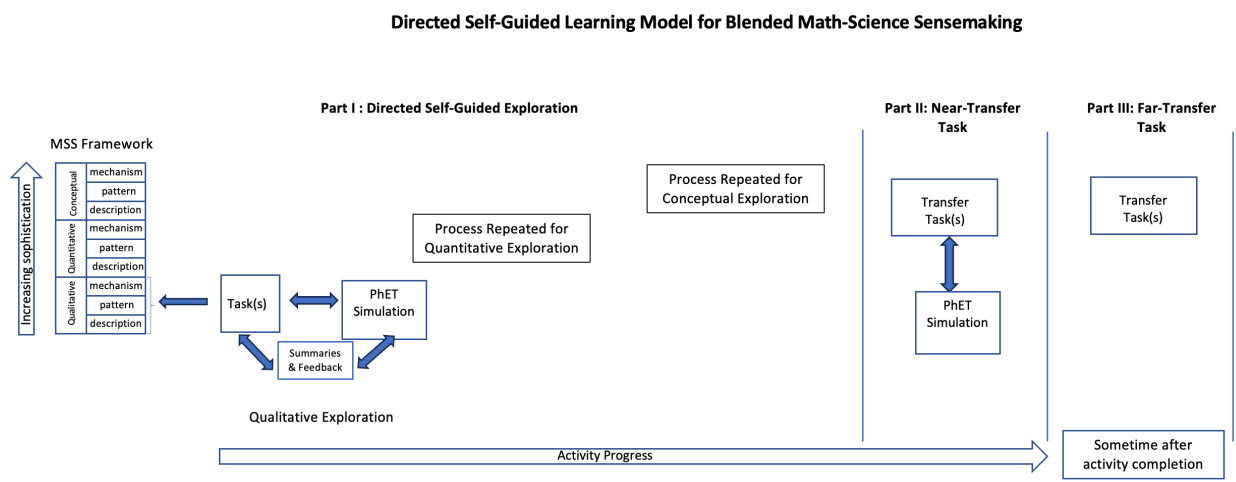

The model represents the structure for MSS learning activities. Each activity task is aligned to specific MSS cognitive framework level (Kaldaras & Wieman, 2023a). Students complete the tasks as they interact with PhET simulations. The tasks are a combination of multiple choice (MC) and short constructed-response (CR) tasks. Students generally start with a pre-assessment question (not shown in Figure 2) asking them to suggest a mathematical formula describing observations based on interacting with the simulation only. The question measures the students' MSS levels before the self-guided activity. After submitting their response to the pre-assessment, students continue to the activity. In the activity, students respond to questions aligned to the MSS level in a manner of increasing sophistication: starting from the lowest (Qualitative) MSS level and proceeding to the highest (Conceptual) level as they progress in the activity. Students respond to questions using the simulation and sometimes additional data provided to them in the form of data tables or graphs. Upon completing the tasks for each MSS framework level (Qualitative, Quantitative, Conceptual), students receive feedback in the form of correct responses and short text-based summaries discussing ideas students need to make sense of the correct responses and proceed to higher MSS level tasks. They are given an opportunity to reflect on their responses before they proceed to higher level MSS tasks. This reflects the guided inquiry approach supported by simulation-based scaffolds, tasks, feedback, and summaries.

The learning model described above reflects *directed* self-guided approach to the learning process, and it is quite different from SDL and SGL learning models described above and shown in Figure 1. In this model, we scaffold the learning process with the sets of questions, feedback, and summaries described, proceeding from less to more sophisticated. The directed guidance follows the MSS cognitive framework (Table 1). This scaffolds the learning process according to developmentally appropriate steps (Bransford, 2000).

Note that MSS framework levels are described in terms of learning performances that reflect specific process-oriented goals that students achieve as they are building better MSS skills (e.g., learning to recognize quantitative patterns and use these patterns to suggest a meth relationship describing observations). Consequently, the tasks in the activities are aligned to specific framework levels and represent specific context-related and process-oriented goals that students attain with the help of the computer simulation to build better, transferable MSS skills. The activity guides students through the *process* of math-science sensemaking emphasizing all the important, cognitively appropriate steps towards developing a math equation, rather than outcome-oriented goal (e.g., develop an equation based on observations without scaffolding). This is valuable for developing deep understanding and transferable skills (Brydges et al., 2009).

## Methods

### General Activity Design Structure

The activity starts with a set of tasks aligned to the lowest MSS level – level 1 "Qualitative". These tasks guide students to use the simulation to answer questions on the variables that affect behavior in the sim (level 1 "Description" MSS sub-level), the qualitative patterns among the variables (level 1 "Pattern" MSS sub-level) and underlying qualitative causal mechanism (level 1 "Mechanism" MSS sub-level). Students submit their responses to these qualitative level questions, and then receive feedback in the form of correct responses to the questions and examples of accurate reasoning. Students are then given the opportunity to reflect on their responses, compare their responses to the correct response, and use the simulation again to help in the reflection process. Next, students are provided with short text-based summaries to help them understand the qualitative observations and introduce successful engagement in MSS. The first summary focuses on introducing and explaining basic math relationships in the sim.

The current study focuses on phenomena involving proportional and quadratic relationships. Below is an example of a summary. Such summaries prepare the students to engage in higher level, level 2 (Quantitative) type of MSS.

*Table 2. Example of summary providing information on fundamental math relationships that can be used to engage in level 2 type of MSS.*

How can we describe these observations mathematically? One possibility is a proportional relationship. Proportional relationships reflecting different types of dependencies are shown below in general form, where "Y" is the dependent variable, "X" is the independent variable and "k" is proportionality constant.

1) $Y = k \times X$ - directly proportional relationship is described by multiplication operation and reflect the type of dependency where the variables (X and Y) increase or decrease together by some proportional amount determined by the constant "k" .

2) $Y = k/X$ - inversely proportional relationship is described by division operation and reflect the type of dependency where "Y" decreases as "X" increases by a certain proportional amount described by proportionality constant "k".

3) $Y = X^2$ ("x" squared) - quadratic proportional relationship reflects that increasing X would result in increasing Y in a quadratic proportion.

4) $Y = Z/X$ - this equation reflects both direct and inverse proportional relationships. Specifically, increasing Z leads to increase in Y because Y is directly proportional to Z (hence, Z in in the numerator), and increasing X leads to decrease in Y because Y is inversely proportional to X (hence, X is in the denominator).

It is important to look at how the numerical values of Y change with respect to X and try to identify specific patterns that can describe that change.

Note that students are given information on fundamental math relationships, but they are the ones who decide how to use this information and interaction with the sim to come up with the equation that best describes their observations. This is the directed self-guided approach.

After working on level 1 questions, students proceed to the next set of questions aligned with level 2 "Quantitative" MSS. These questions guide students to use the sim and sometimes other resources (e.g., data tables, graphs) to notice the numerical values of variables that describe observations in the sim (level 2 "Description" MSS sub-level), the quantitative patterns among the variables (level 2 "Pattern" MSS sub-level) and propose and justify a math relationship based on everything learnt so far. A constructed response question asks the student to provide the equation on their own and justify it, and then they are given a MC question where students are asked to pick the most likely equation and justify it. The reason for two formats is to give students the opportunity to find the equation on their own first (CR), and then provide them with additional scaffolding in the form of provided choices if necessary.

Common alternatives for responses are a) justifying by plugging in the numbers from data tables or sim to arrive at the correct equation, consistent with level 2 "Mechanism" type of MSS (see Table 1) or b) using the summaries and their observations of quantitative patterns in the sim, level 3 "Pattern" type of MSS (see Table 1).

Once students complete tasks associated with Level 2 Quantitative MSS they are again provided with feedback, summaries and opportunity to reflect on their response using the sim. These summaries support learners in translating the general mathematical relationships discussed earlier into a mathematical equation describing the given phenomenon. For example, in the context of the Coulomb's Law sim, the specific inverse quadratic relationships are pointed out between the magnitude of interacting charges and the associated magnitude of electric force between the charges (Table 3). The purpose of the summary is to provide additional supports to ideally help all students attain at least level 3 "Pattern" type of MSS reflected in the ability to relate quantitative patterns to the equation structure. This summary provides a more specific information on how to use patterns discussed in the first summary (Table 2) to engage in MSS with this phenomenon. Note that the summary demonstrates the *process* of MSS for students that leads to noticing inverse quadratic relationship based on observations in the sim. This reflects additional scaffolding of MSS for students aimed at helping them understand how to relate observations to specific math relationships and equation structure, which should help them attain level 3 "Pattern".

Table 3. *Example of summary providing information on relating fundamental math relationships to the observations in the sim.*

You have seen on the previous page that increasing the distance between the charges by some amount causes the magnitude of the associated attractive or repulsive force between the two charges to decrease by that amount squared. You can easily observe that using the simulation below. Set both charges to 1 for simplicity. Then, increase the distance between the two charges by two times, from 2 to 4 cm. You can see that the magnitude of the electric forces decreases from about 22.5 N to 5.6 N or decreases by 4 times. Notice that 4 is the square of 2, which is the amount by which the distance was changed in this case. Similarly, say you now increase the distance by three times compared to the original distance: from 2 cm to 6 cm. You can see that the force changes from 22.5N to about 2.5 N or decreases by 9 times. Notice that 9 is the square of 3, which is the amount by which distance was changed in this case. This pattern holds irrespective of what values for the distance and magnitude of charge you try (feel free to explore this with the sim).

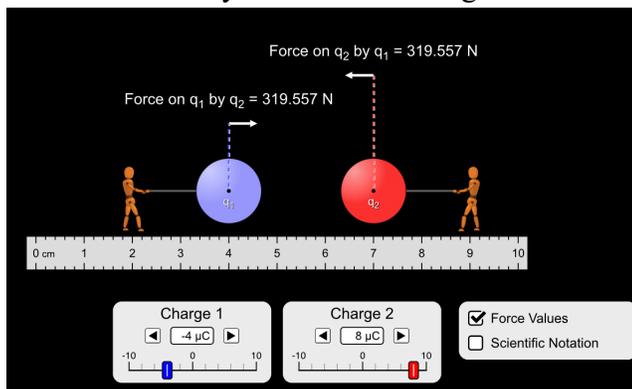

This pattern suggests the magnitude of electric force between the two charges is proportional to the inverse square of the distance between the two charges. Recall from previous summary that an inversely proportional relationship describes situations where as one variable increases, the other variable decreases by some proportional amount: for example, as one variable increases (distance in this case), the other one decreases (magnitude of the electric force between the charges in this case) by the proportional amount equal to the square of the change.

$$\text{Force} = \frac{1}{\text{distance}^2} \times k$$

Upon completion of level 2 "Quantitative" MSS tasks and reflection students proceed to tasks aligned to level 3 "Conceptual". The nature of these tasks and questions can vary, but the goal is to help students to reach level 3 "Pattern", relating quantitative patterns to the equation structure and level 3 "Mechanism", developing an explanation for how the proposed equation structure reflects causal mechanism of the phenomenon under study. We have tried different task formats, including asking students to explicitly relate their proposed equation structure to the simulation behavior; and providing students with three examples of arguments for why the proposed equation makes sense and asking them to choose the best one and explain their choice. The arguments each focused on different strategies for coming up with the equation consistent with level 2 or level 3 of the framework. Example of this task is shown in Table 4.

*Table 4. Example of Level 3 MSS Task. Note: MSS framework level alignment is shown in red and is not part of the original prompt.*

A group of friends completed the activity you just did and arrived at the following equation:

$$Force = \frac{Q1 \times Q2}{distance^2}$$

Each friend offered an explanation for the process described by this equation.
Pick the explanation for the process that is the most accurate and complete in your opinion.

<u>Friend 1</u>: the magnitude of electric force between the two charged objects is primarily determined by the magnitude of charge on each interacting object and the distance between the charges. Larger charge on the interacting objects results in larger electric force, which is reflected in "Q1" and "Q2" being in the numerator of the equation. Smaller distance between the charges results in larger electric force, which is reflected in "distance" being in the denominator of the equation. Since changing the distance between the objects affects the magnitude of the electric force more than changing the magnitude of charge in the interacting objects, the distance variable should be squared as shown in the equation.
*<u>Note</u>: this sample reasoning is alignment to level 2 "Mechanism" type of MSS because the student is providing qualitative justification for the equation structure.*

<u>Friend 2</u>: the magnitude of electric force between the two charged objects is primarily determined by the magnitude of charge on each interacting object and the distance between the charges. The simulation shows that increasing the charge on either object by certain factor (e.g., factor of 2 or 10) increases the electric force between them by the same factor. This suggests a direct linear relationship between the electric force and the amount of charge on each object, "Q1" and "Q2". Hence, "Q1" and "Q2" are in the numerator. Further, the simulation shows that increasing the distance between the charges by a certain factor decreases the magnitude of the associated electric force by the square of that factor. This suggests an inverse squared relationship between the magnitude of electric force and the distance between the two charges. Hence, distance squared is in the denominator.
*<u>Note</u>: this sample reasoning is alignment to level 3 "Pattern" and "Mechansim" type of MSS because the student is providing quantitative pattern-based justification for the equation structure.*

<u>Friend 3</u>: the simulation shows that increasing the charge on either one of the interacting objects and keeping the distance between them unchanged increases the magnitude of the associated electric force. The simulation also shows that increasing the distance between the two objects and keeping the amount of charge on each object unchanged decreases the magnitude of the associated electric force. The numbers for associated variables (Q1, Q2 and distance between the charges) in the simulation suggest that multiplying Q1 by Q2 and then dividing by the distance squared will yield the magnitude of the associated electric force.
*<u>Note</u>: this sample reasoning is alignment to level 2 "Mechanism" type of MSS because the student is providing qualitative and "plugging in the numbers" type of justification for the equation structure.*

Notice that each sample argument focuses on the type of MSS reasoning consistent with either lower level MSS (level 2 "Mechanism") or higher-level target MSS (level 3 "Pattern" and "Mechanism"). Sample arguments reflect the most common research-identified types of reasoning students tend to use to justify the proposed equation. Once students submit their choice and written explanation for the choice, they are provided with a summary giving the argument with the targeted level 3 type MSS reasoning (Friend 2 in the example provided in Table 2) and an explanation for why this is the best justification of the three.

Finally, the last task addresses level 3 "Description" MSS by guiding students to think about the scientific meaning of the variables and constants that are not obvious, but important to include into the equation. In the context of Coulomb's law this would be Coulomb's constant, which depends on the medium in which charges are interacting. The task guides students to understand both the mathematical need to include the Coulomb's law constant into the equation as a proportionality constant, and the scientific meaning of the constant.

To summarize, the activity scaffolds student engagement in the process of MSS by guiding students in exploring qualitative variables, patterns and mechanism (level 1), followed by evaluating quantitative patterns among the relevant variables using data and sim and starting to learn how to use the data to develop an equation (level 2), followed by supporting students in relating the patterns to the equation structure, develop a quantitative causal mechanism, and evaluating and justifying the need for any unobservable variables or constants (level 3).

Once the students complete the tasks associated with level 3 of the framework, they complete a transfer task (see Figure 1) in the same session. The goal of the transfer task is to gauge the degree of short-term transferability of MSS skills focusing on near transfer and some elements of far transfer. The transfer task is a simulation on a different phenomenon with a question asking students to propose a mathematical equation describing the simulation behavior and justify the proposed relationship. There is no scaffolding for this activity. Students can go back to look at the prior summaries and work, but the transfer task involves a totally different context for MSS application.

In the context of MSS, it is important to consider several factors when deciding on whether a given task represent a far or near transfer. Specifically, recall that engagement in MSS is reflected in students being able to integrate two cognitive dimensions: science and math (Zhao & Schuchardt, 2020). Therefore, MSS transfer could be related to transferring MSS skills to a new science context, new math context, or both. Therefore, the transfer task phenomenon can be chosen to reflect any of these transfer types for MSS. For example, a transfer task phenomenon could have a different physical context but has a similar underlying math relationship, such as proportional relationships. This is an example of far transfer for science cognitive dimension, and near transfer for math cognitive dimension. Alternatively, a transfer task phenomenon could have a similar physical context but a different underlying math relationship. This is an example of near transfer for science cognitive dimension, and far transfer for math cognitive dimension. Finally, a transfer task phenomenon could have an entirely different physical context and underlying math relationships, which would be reflective of far transfer on both science and math cognitive dimensions. In this study, the post activity transfer task reflected a different scientific context, so far transfer for the science dimension, and some similar and some different mathematical relationships, so elements of both near and far transfer for the math dimension.

Finally, all students then complete a post assessment at least one week after the activity and transfer task completion. This MSS assessment was developed and validated previously (Kaldaras & Wieman, in preparation). This assessment measures farther and longer-term transfer. The assessment consists of several test lets (3-5 items connected by the same prompt) probing MSS in various novel STEM contexts (simple Chemistry, Physics, and interdisciplinary scenarios) at all MSS framework levels. No interactive simulations are involved. All items are text- based containing all the necessary information to engage in MSS in the form of tables, figures, and graphs. Items also focus on a wider range on underlying math relationships than those covered in the original activity. For example, if the activity only focused on directly proportional relationships, the assessment tasks focus on inversely proportional and quadratic relationships, among others. This design has some elements of near and far transfer on the math dimension and far transfer on the science dimension.

Student learning outcomes are evaluated in several ways. First, each student is evaluated on whether they have developed an accurate math relationship/equation and how they justified the relationship in the original task. We evaluate whether the justification logic is consistent with level 1, 2 or 3 to assign an appropriate MSS level. We use the highest MSS level demonstrated by each student in completing the activity. Second, we evaluate the students' MSS levels demonstrated on the short-term transfer task in the same way. Both evaluations use rubrics designed for the specific scenarios covered in the activity and the transfer task. Examples of such rubrics are shown below (Table 9). Finally, we evaluate the students' performance on the delayed (long-term transfer) assessment using a previously designed rubric (Kaldaras & Wieman, in preparation).

**Activities designed for this study**

There were two activities designed for this study. One activity focused on Physics context and explored the phenomenon of Coulomb's law. The other activity focused on Chemistry context and explored the phenomenon of Heat Capacity. Both activities were designed following the DSGL model described above and shown in Figure 2. Full versions of both activities are provided in the Appendix.

Briefly, the Coulomb's law activity used Coulomb's law simulation (simulation snapshot is shown in Figure 3) to guide students in developing an equation for Coulomb's Law. The target formula states that the magnitude of electric force (F) between the two interacting charges is directly proportional to the magnitude of each of the interacting charges (Q1 and Q2) and inversely proportional to the square of the distance between the two charges, or $F = Q1 \times Q2/distance^2$. The tasks focused the following aspects of the phenomenon:

- identifying the variables that affect the magnitude of electric force (e.g., magnitude of charges, distance between charges) – *Level 1 "Description" MSS*
- identifying the qualitative patterns among the variables (e.g., increasing the magnitude of charges increases the electric force, increasing the distance between the charges decreases the magnitude of force) – *Level 1 "Pattern" MSS*
- developing qualitative causal mechanism for the phenomenon (e.g., electric charges cause attractive or repulsive forces) – *Level 1 "Mechanism" MSS*
- identifying the numerical values of relevant variables and the quantitative patterns among the variables (e.g., increasing the magnitude of one or two of the interacting charges increases force by the same proportional among, decreasing the distance between the charges decreases force by a factor of distance squared) – *Level 2 "Description" and "Pattern" MSS*
- force is caused by interacting charges and depends on the distance between them and the magnitude of the charges. The formula $F = Q1 \times Q2/distance^2$ makes sense when you plug in the associated variable values, but you also need to always multiply by the same number equal to $8.99 \times 10^9 \ N \cdot m^2 / C^2$, which seems to be necessary for making this equation work. – *Level 2 "Mechanism".*

- developing an equation and quantitative causal mechanism for describing observations in the sim (e.g., electric charges cause attractive or repulsive electric forces, and the force depends on the magnitude and distance between charges. Therefore, F= Q1× Q2/distance² which reflects that increasing magnitude of interacting charges causes a directly proportional increase in the force and decreasing the distance between charges causes increase in the force as a function of distance squared. – *Level 2 "Pattern" and "Mechanism".*

Note that we did not have tasks that explicitly focus on Level 3 "Description" reflected in helping student understand the scientific meaning of Coulomb's law constant. This was because the simulation did not support development of this understanding, because it only shows charges interacting in one medium. Thus, for simplicity, we chose to simply provide this information to students.

Further, the Heat Capacity activity used Energy Forms and Changes simulation (simulation snapshot is shown in Figure 4) to guide students in developing an equation for Heat Capacity. The target formula states that the amount of energy, (E) needed to raise the temperature of a given amount of substance by specific number of degrees is directly proportional to the mass (m), specific heat capacity constant (c) and the desired temperature change (ΔT), or E = c × m × ΔT. The students were guided to engage in MSS at all MSS framework sub-levels starting from lowest and counting to the highest ones.

*Figure 3. Coulomb's Law simulation snapshot.*

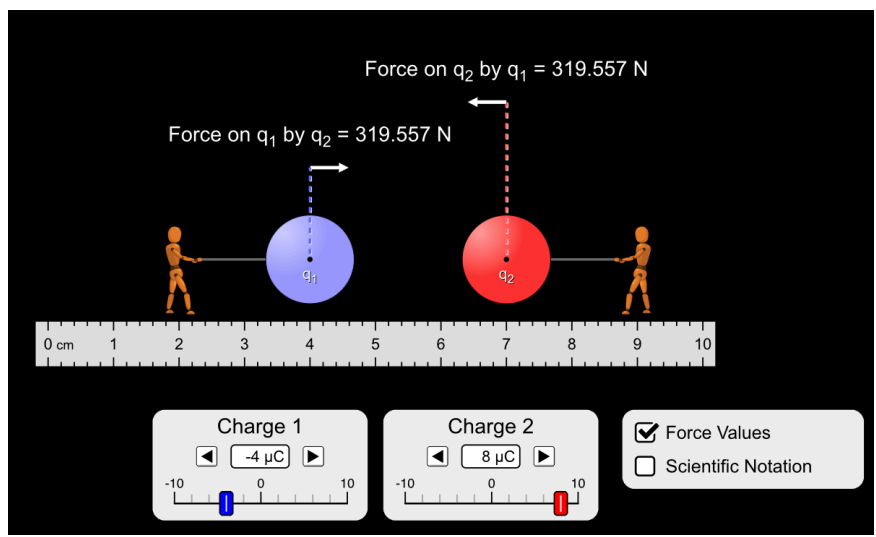

The Heat Capacity tasks focused the following aspects of the phenomenon:

- identifying the variables that affect the energy required to raise the temperature of a substance (e.g., mass, desired temperature change, type of substance reflected in specific heat capacity constant) – *Level 1 "Description" MSS*

- identifying the qualitative patterns among the variables (e.g., increasing the mass of a give substance requires more energy to raise the temperature by the same number of degrees) – *Level 1 "Pattern" MSS*

- developing qualitative causal mechanism for the phenomenon (e.g., heating up a substance causes the temperature of the substance to increase; the energy amount needed to increase the temperature of the substance depends on the type of substance, the mass, and the overall temperature change) – *Level 1 "Mechanism" MSS*

- identifying the numerical values of relevant variables and the quantitative patterns among the variables (e.g., energy needed to raise the temperature of a substance is directly proportional to the mass, specific heat capacity constant and the overall temperature change for the substance) – *Level 2 "Description" and "Pattern" MSS*

- the energy amount needed to increase the temperature of the substance depends on the type of substance, the mass, and the overall temperature change. The formula $E = c \times m \times \Delta T$ makes sense when you plug in the associated variable values, but you need to use the value for variable "c" that is specific to the substance you are studying for making this equation work. – *Level 2 "Mechanism".*

- Substances differ on the amount of energy it takes to raise their temperature. This property is described by specific heat capacity constant included in the equation $E = c \times m \times \Delta T$ – *Level 3 "Description".*

- developing an equation and quantitative causal mechanism for describing observations in the sim (e.g., energy required to increase the temperature of a substance is directly proportional to the mass, specific heat capacity constant and the overall temperature change for the substance, or $E = c \times m \times \Delta T$) – *Level 2 "Pattern" and "Mechanism".*

*Figure 4. Energy Forms and Changes simulation snapshot.*

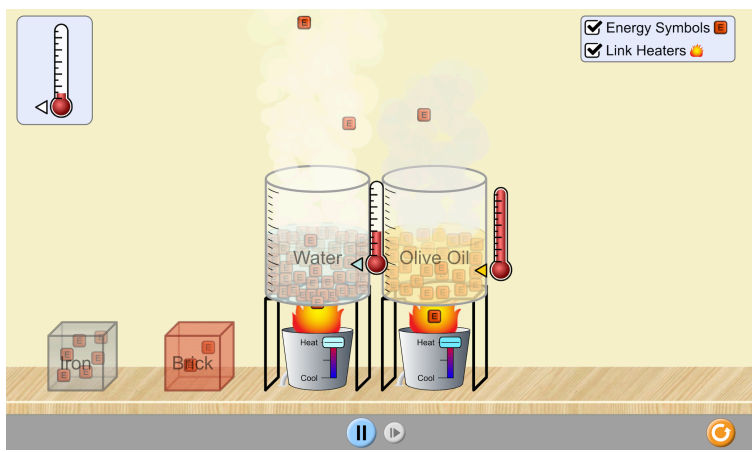

**Student Sample**

We recruited the students by sending out emails to professors teaching introductory STEM courses in minority-serving community colleges and deans of various STEM departments at non-profit community colleges (2-year institutions) from the list provided by Bill and Melinda Gates (BMG) foundation. We specifically targeted institutions with high percentages (50% and above) of enrollment of students from stereotyped racial minority backgrounds ("SRM"). In the email we asked faculty in STEM courses to share information about paid interview opportunities. We have contacted over 50 professors and administrators in over 50 institutions and received a few positive responses indicating the information was shared with students at the corresponding institution. We have also reached out to professional acquaintances in the University of Puerto Rico. As a result, we recruited students from the institutions shown in Table 5.

*Table 5. Educational Institutions attended by student interview volunteers.*

| Institution | Location | N of Students | % SRM enrollment - 3yr avg |
|:---:|:---:|:---:|:---:|
| Long Beach City College | CA | 5 | 70 |
| Bakersfield College | CA | 5 | 71 |
| Santa Fe Community College | NM | 1 | 54 |
| University of Puerto Rico Rio Piedras | PR | 5 | 87 |
| San Jose State University | CA | 6 | 73 |
| Lone Star College | TX | 3 | 65 |
| San Antonio College | TX | 1 | 82 |
| Amarillo College | TX | 1 | 55 |

A total of 126 students from the institutions shown in Table 2 completed the volunteer form that asked them about their academic background and identification with specific minority groups historically marginalized in STEM. "Black" SRM group type refers to dark-skinned people of African descent, no matter their nationality. "Person of Color" SRM group expands the definition of "Black" to include people who are representative of non-White groups such as Latinos, Asians, Native Americans among others. What a respondent prefers comes down to a personal choice and how invested they are into their racial identity. When picking students for interviews, we ensured that students who participated in interviews were freshmen enrolled in introductory Chemistry and/or Physics courses. This ensured that we target the incoming freshmen population from backgrounds historically marginalized in STEM. We recruited a total of 27 students. Each student was compensated with Amazon gift cards for completing the activity and the post assessment. Group-level student demographics are shown in Table 6.

*Table 6. Group-level student Demographics*

| SRM group type | Number of self-identified students |
|---|---|
| LatinX | 15 |
| Person of Color | 10 |
| Indigenous | 1 |
| Black | 4 |
| Women | 13 |
| Identify in two groups | 9 |
| Identify in three groups | 3 |

Individual-level demographics is shown in Table 7.  Academic background including the highest Biology, Physics, Chemistry and Math courses taken for each participant is shown in Table 8.

*Table 7. Demographics background for each participant.*

| Student | POC* | LatinX | Indigenous | Black | Women |
|---------|------|--------|------------|-------|-------|
| 1 | Yes | | | | |
| 2 | Yes | | | | |
| 3 | Yes | | | | |
| 4 | Yes | | | | Yes |
| 5 | | Yes | | | Yes |
| 6 | Yes | | | | |
| 7 | Yes | | | Yes | |
| 8 | | Yes | | | |
| 9 | | Yes | | | Yes |
| 10 | | Yes | | | Yes |
| 11 | | Yes | | | Yes |
| 12 | | Yes | | | Yes |
| 13 | | Yes | | | |
| 14 | Yes | Yes | | | Yes |
| 15 | | Yes | | | |
| 16 | | Yes | | | |
| 17 | | | Yes | | |
| 18 | | Yes | | | Yes |
| 19 | Yes | Yes | | | Yes |
| 20 | Yes | | | | |
| 21 | | Yes | | | |
| 22 | | | | Yes | Yes |
| 23 | | | | Yes | |
| 24 | | Yes | | Yes | Yes |
| 25 | | | | | Yes |
| 26 | | Yes | | | Yes |
| 27 | Yes | | | | |

*POC = person of color*

*Table 8. Academic Background for Each Participant*

| Student | Biology | Physics | Chemistry | Math | Major | English |
|---|---|---|---|---|---|---|
| 1 | Intro Bio | HS Physics | HS Chem. | Algebra I | Film/TV Production | Native |
| 2 | HS Bio | Calc.-Based Phys. II | HS Chem. | Diff. Eq. & linear Algebra | unidentified | Native |
| 3 | Intro Bio | none | HS Chem. | College Stats. | diagnostic Medical Imaging | Native |
| 4 | Intro Bio | none | Prep. College Chem. | Calc I/ College Stats. | Nutrition | Native |
| 5 | Intro Bio | none | Gen. Chem. II | Calculus III | Interdisciplinary Studies | ELL |
| 6 | Intro Bio | none | Gen. Chem. 1 | Calc I/ College Stats. | Occupational Therapy (MS) | Native |
| 7 | HS Bio | none | HS Chem (non AP-AB) | Pre-Calc. | Business Administration | Native |
| 8 | Intro Bio | Calc.-based Physics II | HS Chem | higher than college stats. | Electrical Engineer | Native |
| 9 | Intro Bio | Algebra-based Physics | Gen. Chem II | Calc. 1/ college stats | Biology | Native |
| 10 | Intro Bio | Calc-Based Physics II | Gen Chem II | College Stats. | Biology | ELL |
| 11 | Intro Bio | none | Gen. Chem. II | College Stats. | Biology | Native |
| 12 | Intro Bio | HS Physics | Gen. Chem. II | Calc. 1/ college stats | Biology | ELL |
| 13 | Advanced college Bio | none | Gen. Chem. II | Calc. 1/ college stats | Forensic Science | Native |

| 14 | HS Bio | AP-IB Physics | Gen. Chem. II | Calc I | Mechanical Engineering | Native |
|----|--------|---------------|---------------|--------|------------------------|--------|
| 15 | intro Bio | Calc.-based Physics II | Gen. Chem. I | math above Calc. I | Mechanical Engineering | Native |
| 16 | Advanced Bio courses | none | Gen. Chem. I | pre-calc. | associate in science | Native |
| 17 | Intro College Bio | none | HS Chem. | college stats. | Biology | Native |
| 18 | HS Bio | none | HS Chem. | college stats. | Child and Adolescent Development | Native |
| 19 | HS Bio | HS Physics | HS Chem. | Calc. I | chemical engineering | Native |
| 20 | Intro Bio | Calc.-Based Physics II | Gen. Chem. I | above college stats | Physics | Native |
| 21 | Intro Bio. | none | Gen. Chem. I | pre-calc. | Behavioral Science | Native |
| 22 | HS Bio | HS Phys. | HS Chem | pre-calc. | Kinesiology | Native |
| 23 | Advanced Bio | AP/IB Physics | Gen Chem II | Calc 1 | Biomedical Engineering | Native |
| 24 | Advanced Bio | Calc.-Based Physics I | Gen Chem II | college stats. | Biology | Native |
| 25 | Advanced Bio | Calc.-Based Physics II | Gen Chem II | Calc 1 | unidentified | ELL |
| 26 | Advanced Bio | Algebra-Based Physics I | Gen Chem II | college stats. | Biology | Native |
| 27 | Advanced Bio | Calc.-Based Physics I | Advanced HS Chem | above college stats | Aerospace Engineering | Native |

**Study Design**

The study design diagram is shown in Figure 5. As you can see, the student sample was divided into two groups. One group (12 students) completed the Heat Capacity self-guided activity and was given the transfer task focused on Coulomb's law. The other group completed Coulomb's law self-guided activity and was given a transfer task focused on Heat Capacity. The activity and the transfer task were completed in one session. The reason for this design of ordering was to ensure that students in both groups have opportunity to learn about proportional relationships as part of their self-guided learning activity and receive a transfer task in a different science context focused on proportional relationships similar to those covered in the self-guided activity. Finally, 25 students out of the 27 completed the post MSS assessment between one and four weeks after doing the activity and the transfer task.

*Figure 5. Study Design.*

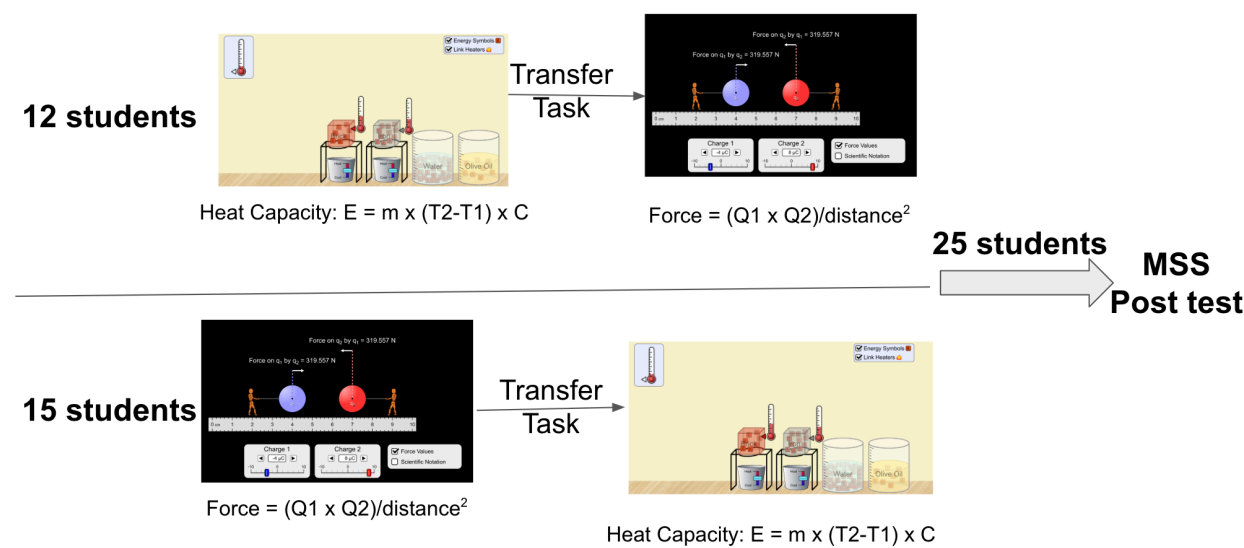

Note that the post activity transfer tasks in this study reflect far transfer in the science dimension for both groups. However, students who complete the Coulomb's law activity explored both directly proportional and inverse squared mathematical relationships, while their transfer task only involved direct proportional relationships so near transfer in math. The situation was reversed for the group doing the heat capacity first, so for them the transfer in the math dimension was farther.

Similarly, the MSS posttest assessment contained two testlets that focused on all of the math relationships covered in both activities (linear and inversely proportional relationships, quadratic relationships) in different STEM contexts. Since students were divided in groups that completed only one of the activities, this assessment represents element of both near and elements of far transfer on the math component for the math relationships that were not part of the activity covered by a given group of students. Therefore, the posttest in this context represents an example of long-term far transfer on the science dimension and a near with elements of far transfer on the math dimension.

**Data Collection**

The activities were administered using the Qualtrics survey tool. All data was collected via a Zoom platform using standard Zoom recording features. Students completed the self-guided activities and transfer tasks as part of the same zoom session that lasted about 1.5 hours. Students were on zoom with a researcher from the project. The researcher shared Qualtrics link with students, explained the self-guided nature of assignment to the students, asked them to share their screen, and recorded all student interactions with the sim and responses to the activity questions. The researcher was present in case any technical issues came up during the session but did not interact with students in any way as they were completing the activity and the transfer task. Students were compensated with $40 Amazon gift wards for completing the activity. All student responses were collected via the Qualtrics platform and later analyzed along with videos of the sessions. The MSS posttest was about 1 hour long and completed in a similar way via the Qualtrics platform. Students were compensated with $20 Amazon gift card for completing the MSS posttest assessment.

**Data Analysis**

We analyzed three data sources, student pre and activity performance, student transfer task performance, and student delayed post-assessment performance.

Student performance on the pre test and the activity and the short-term transfer task was evaluated using previously published general scoring rubric (Kaldaras & Wieman, 2023b). The rubric is shown in Table 9. Notice that it focuses on evaluating two aspects of student response: 1) whether they have proposed an accurate formula; 2) how they justify the proposed formula with respect to MSS framework levels. This rubric allows for accurate MSS level placement of each student response. The task specific examples of student reasoning for each rubric category are provided in the Appendix for both Coulomb's Law and Heat Capacity activities.

The pre assessment measured the students' starting MSS levels. Next, we evaluated how they proposed and justified the math relationships they provided at the end of the learning activity using the same rubric. This gave their final activity MSS level. Finally, we used the same rubric to evaluate student performance on the transfer task.

*Table 9. Scoring Rubric for formative assessments (Kaldaras & Wieman, 2023b).*

| Accurate Formula Provided? | Elements of Student Justification | Cognitive Framework Level |
|---|---|---|
| No | No justification | Level 0 (assumed the student has no idea) |
| Yes | No justification | Cannot be accurately determined |
| No | Lists only variables relevant for describing the phenomenon mathematically | Level 1 (Description) |
| No | Described qualitative relationships among the relevant variables | Level 1 (Patterns) |
| No | Describes qualitative causal mechanism for the phenomenon | Level 1 (Mechanism) |
| No | Described numerical values of the relevant variables | Level 2 (Description) |
| No | Described quantitative patterns among the relevant variables | Level 2 (Pattern) |
| Yes | Justifies the math relationship using numerical values of the variables and/or qualitative causal mechanism or patterns | Level 2 (Mechanism) |
| Yes | Justifies the math relationship by relating the math operations in the equation to quantitative patterns observed in the sim and /or the data. | Level 3 (Pattern) |
| Yes | Provides the causal mechanistic explanation of the equation structure | Level 3 (Mechanism) |

Finally, student performance on the delayed MSS assessment was evaluated on each testlet separately, and an average performance between the two testlets was used as the final performance measure on the delayed MSS assessment. For example, if students demonstrated level 3 type of MSS on one of the testlets, and level 2 on the other, their final MSS assessment score would be 2.5. We evaluated student performance on each testlet using previously designed and validated task specific rubrics (Kaldaras & Wieman, in preparation) grounded in the rubric similar to the one shown in Table 9.

## Results

Tables 10 and 11 shows student MSS level for each of the analyzed data sources (pre assessment, activity, transfer task, delayed post-test) for Heat Capacity and Coulomb's Law respectively. We also include information on the resources student used most during activity. We will further discuss results for each activity in more detail.

### Coulomb's Law Self-Guided Activity

*Table 10. MSS level placement for Heat Capacity activity and Coulomb's Law transfer task.*

*Note: darker cell colors indicate higher MSS framework level.*

| Student | Pre-Activity | Activity | Transfer Task | Post-test | Resources used most in activity |
|---|---|---|---|---|---|
| 1 | 1 (Pattern) | 1 (Pattern) | 1 (Pattern) | 2 (Pattern) | sim |
| 2 | 1 (Pattern) | 3 (Mechanism) | 3 (Mechanism) | 3 (Mechanism) | Sim + summaries |
| 3 | 1 (Pattern) | 3 (Mechanism) | 3 (Mechanism) | 2 (Mechanism) | Sim + summaries |
| 4 | 1 (Pattern) | 2 (Mechanism) | 3 (Mechanism) | 2.5 | Sim + data |
| 5 | 1 (Pattern) | 2 (Mechanism) | 2 (Pattern) | 2 (Mechanism) | Sim + summaries |
| 6 | 2 (Pattern) | 3 (Mechanism) | 3 (Mechanism) | 2.5 | Sim + summaries |
| 7 | 2 (Pattern) | 2 (Pattern) | 2 (Pattern) | 2.5 | Sim + summaries |
| 8 | 3 (Mechanism) | 3 (Mechanism) | 3 (Mechanism) | 3 (Mechanism) | sim |
| 9 | 2 (Pattern) | 3 (Mechanism) | 2 (Mechanism) | 2.5 | Sim + summaries |
| 10 | 1 (Pattern) | 3 (Mechanism) | 2 (Mechanism) | 3 (Mechanism) | Sim + summaries |
| 11 | 2 (Pattern) | 3 (Mechanism) | 2 (Mechanism) | 2 (Mechanism) | Sim + summaries |
| 12 | 2 (Pattern) | 3 (Mechanism) | 3 (Mechanism) | Not taken | Not Taken |

Notice that for Heat Capacity activity most students (except for student 8) start with no accurate equation and a varying degree of quantitative description of the phenomenon. For example, level 1 "Pattern" MSS demonstrated by students 1-5 and 10 reflects qualitative pattern identification and stating for example that "supplying more heat results in higher temperature for the substance".  Similarly, level 2 "Pattern" MSS demonstrated by students 6, 7, 9, 11 and 12 reflects quantitative pattern identification and stating, for example, that "doubling the number of energy cubes added to the substance during heating makes temperature increase by two". This pattern can be determined based on the sim. Only one student (student 8) proposed an accurate formula and justification consistent with level 3 "Mechanism" prior to activity. In their justification, student 8 stated that "energy needed is directly proportional to the change in temperature, mass and specific heat capacity constant". Notice that student 8 demonstrate level 3 MSS across all tasks, which is consistent with our prior findings on transferability of MSS skills across contexts (Kaldaras & Wieman, 2023a).

Most notably, a large fraction of the students (8 out of 12) attained the highest possible MSS level upon activity completion- level 3 (Mechanism). Two students (students 4 and 5) reached level 2 mechanism.  They proposed the correct equation and used the strategy of plugging in the numbers to justify their equation. An example of this justification would be saying that "if I multiply everything together – mass, specific heat capacity and total temperature change- that would give me the correct value for Energy, so E== c × m × ΔT must be correct". Finally, students 1 and 7 remained at their low pre-activity MSS level throughout the activity assessment and transfer task, being unable to ever come up with the correct mathematical equation.

In terms of immediate transfer, four out of seven students who attained the highest MSS level upon activity completion demonstrated the same highest level on the transfer task. The other three students used the strategy of plugging in the numbers to justify the Coulomb's Law equation, which is indicative of level 2 "Mechanism" MSS. In the prior work we found the logic of plugging in the numbers to be very persistent among learners (Kaldaras & Wieman, in preparation), so it is not surprising to see some students use this strategy. Student 5 struggled to come up with an equation for Coulomb's law but did identify quantitative patterns by stating that "force is directly proportional to the amount of charge". Finally, student 4 was notable in that they demonstrated higher level MSS on the transfer task than on the activity task.

The long-term transfer is shown in the post-test column MSS level. The value of 2.5 indicates that students demonstrated level 2 "Mechanism" on one of the testlets and level 3 "Mechanism" on the other testlet. Notice that all students demonstrated higher level MSS on post assessment compared to their pre activity MSS level, indicating long-term transferability of learned MSS skills. Further, notice that most students (7 out of 12) demonstrated the highest possible MSS level (level 3 "Mechanism") on at least one post assessment testlet. Three students demonstrated highest level MSS on both testlets.

**Heat Capacity Self-Guided Activity**

*Table 11. MSS level placement for Coulomb's Law activity and Heat Capacity transfer task.*

| Student | Pre-Activity | Activity | Transfer Task | Post-test | Resources used most in activity |
|---|---|---|---|---|---|
| 13 | 1 (Pattern) | 3 (Mechanism) | 3 (Mechanism) | 3 (Mechanism) | summaries |
| 14 | 1 (Pattern) | 3 (Mechanism) | 3 (Mechanism) | 3 (Mechanism) | Sim + summaries |
| 15 | 1 (Pattern) | 3 (Mechanism) | 3 (Mechanism) | 2.5 | Sim + summaries |
| 16 | 1 (Description) | 2 (Mechanism) | 2 (Mechanism) | Not taken | Sim + summaries |
| 17 | 1 (Description) | 3 (Mechanism) | 3 (Mechanism) | 2 (Mechanism) | Sim + summaries |
| 18 | 1 (Description) | 2 (Mechanism) | 3 (Mechanism) | 2 (Mechanism) | Sim + summaries |
| 19 | 1 (Description) | 3 (Mechanism) | 3 (Mechanism) | 2.5 | Sim + summaries |
| 20 | 1 (Pattern) | 3 (Mechanism) | 3 (Mechanism) | 3 (Mechanism) | Sim + summaries |
| 21 | 1 (Description) | 2 (Pattern) | 1 (Pattern) | 2 (Mechanism) | Sim + summaries |
| 22 | 1 (Description) | 1 (Pattern) | 1 (Pattern) | 1 (Pattern) | sim |
| 23 | 3 (Mechanism) | 3 (Mechanism) | 3 (Mechanism) | 3 (Mechanism) | Sim + summaries |
| 24 | 2 (Mechanism) | 3 (Mechanism) | 3 (Mechanism) | 3 (Mechanism) | Sim + summaries |
| 25 | 3 (Mechanism) | 3 (Mechanism) | 3 (Mechanism) | 3 (Mechanism) | Sim + summaries |
| 26 | 1 (Description) | 3 (Mechanism) | 3 (Mechanism) | 2.5 | Sim + summaries |
| 27 | 1 (Description) | 3 (Mechanism) | 3 (Mechanism) | 3 (Mechansim) | Sim + summaries |

As in the first group, on the pre assessment, most (13/15) students in the Coulombs law activity group showed only level 1 MSS (e.g., charge and distance should be accounted for when calculating the force) either pattern or description. Two (23 and 25) students started at level 3 mechanism, while one started at level two, and all three then remained at level 3 mechanism for all activities.

Upon activity completion most students (9 out of 15) attained the highest possible MSS level -level 3 "Mechanism". These students also demonstrated this MSS level on the transfer task indicating high degree of short-term transfer.  Similarly, 8 out of thee 9 students also demonstrated level 3 MSS on at least one post-assessment testlet, and five of these students demonstrated level 3 on both testlets. Only one student (# 22) did not proceed beyond level 1 on all tasks and a second only reached level 2.  These findings indicate a substantial amount of learning overall from the activity and a high degree of short and long term transfer. The possible reasons for why this group scored higher on the transfer task than the first group is discussed below.

**Resources Used Across the Two Activities**

The simulations and the summaries were the most commonly used resources for both activities as shown in Tables 10 and 11. While we also provided data tables at various points during activities, students tended to rely less on the data tables. This finding is consistent with our prior work where we discovered students from stereotyped racial minority backgrounds (SRM) using simulation more and in more productive ways than data tables to engage in MSS (Kaldaras & Wieman, in preparation).

## Discussion

In this study we introduced a model for directed self-guided learning of MSS skills. The model (Figure 2) is grounded in a previously validated cognitive framework describing increasingly sophisticated ways of engaging in MSS as students are developing a mathematical description of observed phenomena. External scaffolds guide learners to develop a mathematical description of phenomena they are exploring using a PhET sim.  This reflects the model of directed self-guided learning emphasizing process-oriented goals of engaging in MSS, which helps foster transfer (Brydges, 2009). We tested the model with a voluntary sample of first-year college students recruited from various minority-serving institutions across the US and it's territories. To test this model, we designed directed self-guided learning activities focused on MSS for two different contexts: Coulomb's law for Physics and Heat Capacity for Chemistry. Out of a total student sample of 27 students, 12 completed Heat capacity activity and 15 completed Coulomb's law activity.

The results indicate that most students start with the lowest MSS framework level – level 1 (Qualitative). This is consistent with our prior findings indicating that students don't naturally engage in MSS and need additional supports to start developing these skills (Kaldaras & Wieman, 2023a). Three students came in demonstrating the highest MSS level and demonstrated that same level across all activities. This is consistent with our prior findings that students who attain level 3 (Mechanism) type of MSS tend to demonstrate it across STEM contexts (Kaldaras & Wieman, 2023a, b). Notice that these students also had taken more advanced Math and majored in STEM-related areas (Table 8) suggesting that MSS proficiency is related to prior Math and Science preparation.

Most notably, nearly all students improved their MSS level as a result of the learning activity, as shown in Tables 10 and 11. Specifically, most students in both activities attained level 3 (Mechanism) level of MSS indicating that they were able to develop an accurate equation and justify it by relating the quantitative patterns to the equation structure. However, there was still a considerable number of students who used the strategy of "plugging in the numbers" to justify their proposed mathematical relationship reflects in level 2 "Mechanism" type of MSS. This is consistent with our prior work indicating that this is a persistent strategy for all students across STEM, possibly due to the majority of STEM courses being largely focused on algorithmic approaches to using formulas (Kaldaras & Wieman, in preparation).

These findings indicate that activities designed following the directed self-guided approach shown in Figure 2 are helpful in fostering MSS proficiency development among learners from backgrounds historically marginalized in STEM. However, some students need additional opportunities to engage in self-guided learning of MSS in this manner to ensure that they move beyond level 2 (Mechanism) to attain level 3 type reasoning across STEM contexts. Additional studies with larger number of students and more activities are needed to further confirm this finding.

Further, we studied the transferability of MSS skills upon activity completion by evaluating student performance on transfer tasks completed immediately after the activity (short-term transfer), and on the post-assessment completed at least 1 week after the activity (long-term transfer). All transfer tasks had two dimensions of transfer, a math and a science dimension. In terms of science and math dimensions of transfer all tasks reflected the following features: a) far transfer on the science dimension of MSS because they focused on a different phenomenon than that covered in the activity; b) near and elements of far transfer on the math dimension of MSS because they focused on the same and in some cases slightly different math relationships than those covered in the activity.

Results show considerable short-term transfer of MSS skills for both groups of students. Notably, students who completed Heat Capacity activity (Table 10) show less transfer at level 3 (Mechanism) level compared to students who completed Coulomb's law activity. There could be several explanations for this finding. First, the Heat Capacity simulation is less quantitative compared to Coulomb's Law. Specifically, the sim does not support student exploration of quantitative patterns very well but does show general qualitative relationships between amount of heat supplied to substance, and overall temperature change depending on the substance. To overcome this limitation of the sim, we provided students additional data tables reflecting exact numerical relationships between the relevant variables. However, students tend to rely more on simulations, and it might be harder for them to connect data tables to the sim to effectively engage in identifying quantitative patterns. (Kaldaras & Wieman, 2023b). In contrast, the Coulomb's law simulation allows direct exploration of quantitative patterns between the magnitude of electric force, magnitude of charges and distance between charges. Therefore, the fact that heat capacity sim limits engagement in this critical type of MSS necessary for engaging in level 3 (Mechanism) might influence transferability of MSS skills at the highest framework level for students who did this activity. An alternative explanation might be that the Heat Capacity activity involved less complex mathematical relationships than Coulomb's law activity. Therefore, applying level 3 (Mechanism) MSS reasoning on the transfer task with its more complex math relationship would have been challenging for students who did the Heat Capacity activity. These findings suggest that while the directed self-guided learning approach introduced in this study helps most students develop higher MSS proficiency, additional studies are needed to better understand how simulation features as well as math and science context affect transferability of MSS skills across STEM fields.

Finally, we see that the long-term transfer is also considerable for both activities as reflected in student performance on the post assessment. Most students performed much better on the delayed post-assessment compared to their pre-activity level. Nearly all students were able to develop an accurate math relationship on both testlets and justify the relationship using either the logic of plugging in the numbers (level 2 "Mechanism") or relating quantitative patterns to equation structure (level 3 "Mechanism"). These results suggest that completing just one directed self-guided learning activity following our approach helps students develop transferable MSS skills, which answers our RQ1, and the degree of transfer is considerable for both short term and long-term transfer cases, which answers RQ2. However, additional research is needed to better understand the dependence of transferability of MSS skills on learners' familiarity with the specific science and math dimensions. Further, it is important to develop multiple opportunities for students to practice MSS engagement to ensure that they consistently apply level 3 type reasoning in a range of contexts. This will deepen students' STEM understanding. Future work in this direction will focus on designing additional learning experiences for students from diverse backgrounds aimed at fostering MSS across STEM contexts following directed self-guided model introduced in this study and investigate how these learning experiences will carry over to student performance in STEM courses.

Further, note that two students in this study stayed at low MSS levels throughout the activities (students 1 and 22). Student 22 said they were tired and was not very engaged in the activity, seemingly only focused on rapid completion, but student 1 is more instructive. Upon activity completion, student 1 shared that their poor math understanding prevented them from performing better in the activity tasks, but that they enjoyed the activity. It was obvious that this student struggled to understand fundamental math ideas related to variables, interpreting tables, and noticing relationships between variables. Note that this students' highest math taken is Algebra 1 and they are a non-STEM major. Therefore, we believe there are additional challenges to supporting MSS among learners with very low levels of prior math preparation. The fact that this student found the activities interesting and engaging is encouraging, however.

Finally, two students performed better on transfer task or MSS post assessment compared to the activity (students 7 and 21). While it is a small number of students, these findings hint at some MSS skills developing with some delay after a learning activity. This calls for additional studies on long-term transferability of MSS skills across STEM contexts and their effect of STEM success among diverse learners.

In conclusion, this study introduced a directed self-guided approach to learning MSS skills. We tested this approach with learners from backgrounds historically marginalized in STEM. We discovered that activities designed following this approach are quite successful in helping learners develop transferable MSS skills across STEM contexts. Additional studies are needed to better understand the factors that influence MSS transfer and the degree of transferability of MSS skills across STEM. A limitation of this study was that we had a relatively small number of participants and they spanned many different academic and social backgrounds. Although we see clear trends, a larger sample with a finer differentiation of the characteristics of the sample population would be illuminating.